# Improving Maximal Safe Brain Tumor Resection with Photoacoustic Remote Sensing Microscopy

*Benjamin R. Ecclestone [1], Kevan Bell [1,2], Saad Abbasi [1], Deepak Dinakaran [2,5], Frank K.H. van Landeghem [4], John R. Mackey [2,5], Paul Fieguth [3], Parsin Haji Reza [1]\**

1. PhotoMedicine Labs, Department of System Design Engineering, University of Waterloo, 200 University Ave W, Waterloo, ON, N2L 3G1, Canada
2. IllumiSonics Inc, Department of System Design Engineering, University of Waterloo, 200 University Ave W, Waterloo, ON N2L 3G1, Canada
3. Vision and Image Processing Lab, Department of System Design Engineering, University of Waterloo, 200 University Ave W, Waterloo, ON N2L 3G1, Canada
4. Department of Laboratory Medicine and Pathology, University of Alberta, 116 St & 85 Ave, Edmonton, Alberta, T6G 2B7, Canada
5. Cross Cancer Institute, Department of Oncology, University of Alberta, 116 St & 85 Ave, Edmonton, Alberta, T6G 2V1, Canada

\*Email: phajireza@uwaterloo.ca

**Abstract**

Malignant brain tumors are among the deadliest neoplasms with the lowest survival rates of any cancer type. In considering surgical tumor resection, suboptimal extent of resection is linked to poor clinical outcomes and lower overall survival rates. Currently available tools for intraoperative histopathological assessment require an average of 20 minutes processing and are of limited diagnostic quality for guiding surgeries. Consequently, there is an unaddressed need for a rapid imaging technique to guide maximal resection of brain tumors. Working towards this goal, presented here is an all optical non-contact label-free reflection mode photoacoustic remote sensing (PARS) microscope. By using a tunable excitation laser, PARS takes advantage of the endogenous optical absorption peaks of DNA and cytoplasm to achieve virtual contrast analogous to standard hematoxylin and eosin (H&E) staining. In conjunction, a fast 266 nm excitation is used to generate large grossing scans and rapidly assess small fields in real-time with hematoxylin-like contrast. Images obtained using this technique show comparable quality and contrast to the current standard for histopathological assessment of brain tissues. Using the proposed method, rapid, high-throughput, histological-like imaging was achieved in unstained brain tissues, indicating PARS' utility for intraoperative guidance to improve extent of surgical resection.

# 1      Introduction

With some of the lowest median survival rates of any type of neoplasm, malignant brain tumors are among the deadliest of diseases[1]. The current standard of care for brain tumors is surgical resection, as treatment options are limited by the potential for damage to critical functions of brain tissue, and modest sensitivity to cancer drugs and radiation[2]. During resection, the surgeon must carefully decide what tissue to remove such that a maximal extent of tumor resection is achieved without significant neurologic damage to optimize patient outcomes[3,4]. In the case of diffusely infiltrating gliomas, the most common primary brain tumor, maximal safe resections confer a significant increase in survival time for patients[3,4]. Hence, increasing extent of resection is regarded as an essential factor in improving glioma patient outcomes[3,4].

Resection surgeries are guided by preoperative computed tomography (CT) and intraoperative magnetic resonance imaging (MRI), along with intraoperative neuropathologic analysis[5-7]. While MRI rapidly provides spatial information and reveals dense tumor regions, histopathology is used for identifying microscopic tissue properties, differentiating gliosis vs. healthy tissue, white vs. gray matter and necrotic vs. living tissues. Currently, frozen pathology is the main histopathological technique used intraoperatively. However, frozen histology can only be performed on a macroscopically representative subset of resected tissue, and each such sample requires an average of 20 minutes of processing[8,9]. Additionally, this method is prone to tissue distortion and difficulties with interpretation compared to formalin-fixed paraffin-embedded preparations, thereby providing suboptimal information to the surgeon[10].

In an effort to reduce reliance on suboptimal neuropathology, recent works have explored supplementing traditional MRI with methods such as fluid-attenuated inversion recovery (FLAIR) MRI[11]. While FLAIR successfully highlights areas of diffusely infiltrating tumor, it also captures demyelination, cerebral edema, and surgery-related injury potentially driving excess resection and unnecessary neurological damage to patients[11]. Furthermore, this method is slow, and like standard MRI lacks microscopic precision, contrast and tissue specificity[11,12]. Outside MRI, some researchers have explored tumor identification using *in-situ* fluorescence microscopy[12]. However, these fluorescence methods are only effective in identifying high grade gliomas, are prone to error from obscuration in the surgical space, depend on exogenous dyes and lack discrimination between regions of tissue[13]. Consequently, there is no substitute for intraoperative histopathology when guiding resection surgeries. Hence, there remains an urgent need for a rapid interoperative *in-situ* histopathological imaging technique to aid in safely and maximally resecting brain tumors.

One modality which has shown promise as a label-free histopathological imaging technique is optical resolution photoacoustic microscopy (OR-PAM), a hybrid optomechanical modality which visualizes endogenous optical absorption contrast. In OR-PAM, a pulsed excitation laser is used to generate acoustic pressure waves which are then observed with an acoustically coupled ultrasound transducer[14,15]. By targeting absorption peaks characteristic of specific biomolecules, OR-PAM has demonstrated label-free microscopy of subcellular tissue features[16]. Using UV excitation to visualize DNA, recent works have provided exemplary histology like imaging in human breast tissue and murine brain tissues[14,15,16]. Within brain tissues, OR-PAM has provided rapid histological imaging of entire brain tissue slices using a multifocal UV excitation[15]. However, these UV based OR-PAM systems require physical contact to form images. Within the neurosurgical operating field, physical coupling is infeasible due to the tight spatial constraints, sensitive tissues, and risk of infection. Therefore, while UV-PAM has provided excellent histology-like results label free, there remains a need for an all-optical reflection-mode apparatus more appropriate for in-situ applications.

Recently our team, for the first time, utilized photoacoustic remote sensing (PARS) microscopy to demonstrate nuclear and cellular structure imaging in unstained human tissue preparations[17,18]. PARS operates by directing a nanosecond scale excitation pulse into the sample. As the target absorbs optical energy from the excitation laser it undergoes thermo-elastic expansion, resulting in nanosecond scale modulations in the local refractive index which can be read as back-reflected intensity variations of a second co-focused continuous-wave detection laser[19,20]. These intensity modulations are then proportional to the absorbed excitation energy[19,20]. In this way, the PARS imaging modality visualizes optical absorption contrast in biological samples with an all-optical label-free non-interferometric reflection-mode architecture. The ability to work in reflection mode, rather than requiring light transmission through the target, permits imaging unprocessed thick (>1 mm) tissue samples. In this work we present a multiwavelength rapid acquisition PARS system designed to directly emulate the contrast of histopathological staining in human brain tissues, specifically the widely used hematoxylin and eosin (H&E) staining.

## 2    Methods:

### 2.1    *System Architecture*

The system architecture is outlined in Figure 1. The 266 nm excitation source was a 10 mW, 20 kHz, 0.6 ns pulsed laser (SNU-20F-10x, Teem Photonics). The tunable source was an Ekspla NT210, a 3ns pulsed 1 kHz laser. Excitation laser outputs were spatially filtered prior to combination with the detection beam using a 25 µm pinhole (P25C, Thorlabs) and 50 µm pinholes (P50C, Thorlabs) for the 266 nm and tunable excitation respectively.  Spatial filtering is required as the selected excitation lasers produce astigmatic and highly elliptical beams, inappropriate for near diffraction limited focusing. The detection laser was a 1310 nm superluminescent diode (S5FC1018P. Thorlabs). Horizontally polarized detection passes through the polarizing beam splitter (CCM1-PBS254, Thorlabs) and is converted to a circular polarization state using a quarter wave plate (WPQ10M-1310, Thorlabs). The combined excitation and detection are passed through a 2D galvanometer mirror optical scanning system (GVS412, Thorlabs), then co-focused onto the sample with a 0.3 numerical-aperture 15x reflective objective lens (LMM-15X-UVV, Thorlabs). Circularly polarized back-reflected detection beam is shifted to a vertical polarization state by the quarter wave plate. The vertically polarized detection is directed through a long-pass filter (DMLP1000, Thorlab.), then focused with a condenser lens (ACL25416U, Thorlabs) onto an InGaS photodiode (PDB425C-AC, Thorlabs). The analogue output of the photodiode is then interpreted by a 14-bit digitizer (Gage Applied).

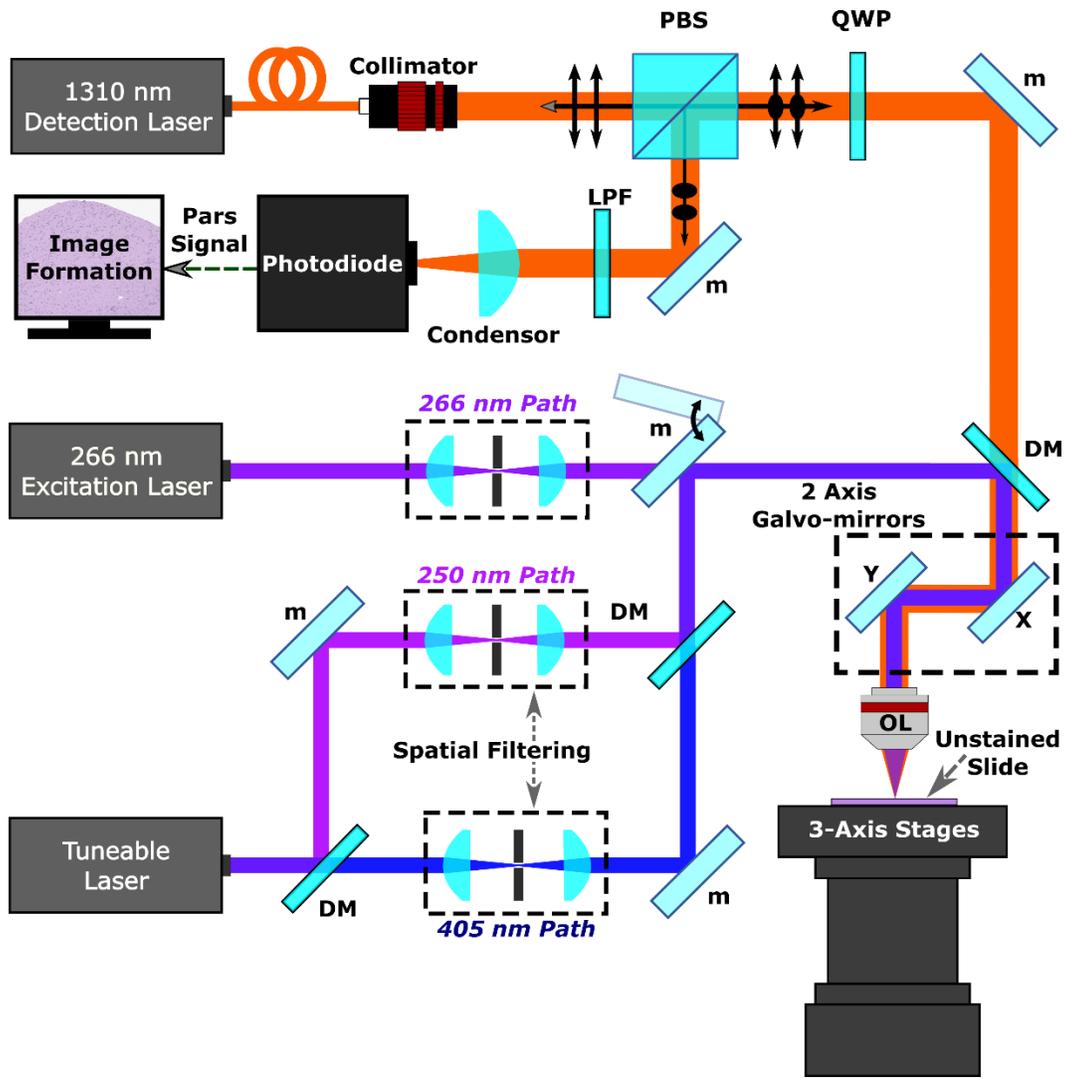

**Figure 1:** Simplified Schematic of the PARS system. Component labels are defined as follows: polarizing beam splitter (PBS), long-pass filter (LPF), quarter-wave plate (QWP), objective lens (OL), dichroic mirror (DM), fold mirror (m).

*2.2   Image Reconstruction*

Generation of a point scanned image was performed with two methods, optical scanning with galvo-mirrors and mechanical scanning of the sample. In each case, upon input of an excitation pulse, the relative location of the interrogation in *x* and *y*, and a short sample of photodiode output was captured by a digitizer. To determine the PARS amplitude, the upper and lower envelopes of the photodiode signal were extracted and used to calculate the absolute magnitude. For the mirror scans, at each PARS excitation the x and y mirror angles were recorded and used to determine the interrogation location. The scattered PARS data was then fitted to a cartesian grid using nearest neighbor search. Linear interpolation was applied across the fitted data to determine the appropriate signal for each image pixel. For the mechanical scans, the stage speed was adjusted to generate equidistant lateral spacing between interrogation points, directly generating a grid of PARS interrogations. For each single wavelength a colormap was applied to this gridded data to form a completed image. To generate the false color H&E images utilizing both 250 and 420 nm data, further processing was applied to the final data sets. First, segmentation was used to determine background from tissue based on nuclear signal density. Second, for tissue regions, pixel values were calculated based on a weighted mixing function between the 250 nm and 420 nm data.

## 2.3 *Sample Preparation*

In this study, a variety of formalin fixed paraffin embedded (FFPE) brain tissue samples were selected for testing. From each FFPE tissue block, two adjacent 4 µm sections were cut and fixed to slides. One of these slides underwent traditional histopathological staining and was subsequently imaged using a conventional bright field microscope. Areas of interest for the histological diagnosis of brain tumors were then identified to test the functionality of the PARS system. The adjacent unstained slices were imaged using the PARS microscope such that a direct comparison could be made between PARS and the current gold standard for histopathological imaging. Formalin fixed paraffin embedded brain tissue samples were obtained under protocols approved by Research Ethics Board of Alberta (Protocol ID: HREBA.CC-18-0277) and University of Waterloo Health Research Ethics Committee (Humans: #40275). The ethics committees waived the requirement for patient consent as the selected samples were archival tissue no longer required for diagnostic purposes, and no patient identifiers were provided to the researchers. All experiments were performed in accordance with the requirements of the Government of Canada, Panel on Research Ethics Guidelines.

## 3      Results and Discussion

In H&E staining, the hematoxylin dye stains cell nuclei while eosin dye marks cellular structure. In order to mimic the contrast of H&E and generate false-color PARS images, a 1 kHz tunable excitation laser was used to target the same biomolecules highlighted in H&E staining. Two excitation wavelengths were selected to conduct PARS imaging. The first wavelength, 250 nm, was selected to provide absorption contrast akin to hematoxylin staining of the nucleus by targeting the UV absorption peak of DNA[16]. The second wavelength, 420 nm, was selected to generate eosin-like cellular structure contrast by targeting cytochromes, a group of extra-nuclear and mitochondrial proteins found in eukaryotic cells[21]. To compensate for limitations imposed by the slow repetition rate of the tunable laser, an additional fast 21 kHz 266 nm excitation laser was added.

Considering the tunable excitation laser, recovering a 1 mm x 1 mm image with 900 nm lateral sampling steps requires 22 minutes per wavelength. Though these scans are limited by the slow repetition rate of the tunable laser, the dual wavelength contrast provides the first true PARS simulated H&E images. Assessing the individual wavelengths, the 250 nm excitation provided $580 \pm 20$ nm spatial resolution, while the 420 nm excitation provided $590 \pm 20$ nm resolution. Depending on factors such as the tissue structure, excitation beam quality, and objective lens, the excitation pulse energies ranged from 0.9nJ to 20 nJ . Though the presented specifications are not ideal for clinical use, this study validates the efficacy of the proposed PARS system in recovering true simulated H&E contrast using a non-contact and label-free microscope in unstained brain tissues. Moving forwards, many non-technical improvements may be implemented such as incorporating faster, shorter pulsed lasers which would drastically reduce the excitation energy required while increasing the imaging speed.

Currently, to mitigate speed limitations, the faster 266nm excitation provides rapid scanning capabilities. Images captured with the 266 nm laser used 0.9 to 20 nJ excitation pulses and delivered 2.5 ± 0.4 µm spatial resolution. Using the 266 nm excitation a 900 nm spatial sampled 1 mm x 1 mm area can be captured in under 1.5 minutes, compared to the 22 minutes required by the tunable laser. In addition to providing rapid grossing scans, smaller fields may be assessed rapidly with optical scanning. In the real-time mode, a 175 µm by 175 µm image, composed of 100,000 interrogation points is captured in 5 seconds, equivalent to a 0.2 fps frame rate. As with the tunable laser, the 266 nm excitation is still limited by repetition rate. While the 266 nm source is substantially faster than the tunable laser, the imaging speed could be improved by a factor of 50 (a frame rate of 10 FPS) using a 1MHz excitation. Moving forwards, a faster laser will be implemented substantially increasing the imaging area, and real-time frame rate.

Starting with the rapid acquisition system, the 266 nm laser was used to quickly image the unstained tissue samples with optical scanning. In this imaging mode, bulk tissue properties such as internuclear distance, nuclear density, cellular composition and organization can be observed, enabling differentiation of necrotic, or neoplastic tissue from healthy tissue. Figure 2a shows a standard brightfield image of solid tumor glioblastoma with microvascular proliferations outlined in red, where the tumor spread extends into the region not resected by the surgeon. In the PARS image from the adjacent unstained section, Figure 2b, the same tumor margins and microvascular proliferations can be identified. Using this real-time capable PARS system, diagnostic properties are recovered from unstained tissues. This represents a critical step towards intraoperatively providing near instantaneous histopathological assessment.

In addition to real-time capabilities, the 266 nm laser's relatively high repetition rate was leveraged to image large FOVs. The ability to quickly preview large regions of tissue is essential in histological analysis. Pathologists may be able to then use these images to scan a large field of tissue and select clinically relevant regions where further higher resolution imaging should be performed. Figure 2c shows a brightfield H&E image of malignant tissue where the blue outline and stars indicate necrotic regions (lacking viable tumor cells with characteristic pseudopalisades) within this region a thrombotic vessel is outlined in purple. Bordering this, the red outline indicates solid tumoral tissue, and microvascular proliferations. In the PARS image taken from the adjacent unstained tissue section, similar bulk tissue properties, including the location of the thrombotic vessel can be observed (Figure 2d). In the PARS image both the necrotic regions and the cell-dense solid tumor are indicated by the same boundaries as in the H&E image. The wide area and real time scanning functions presented here generate images where diagnostic characteristics can be rapidly recovered, allowing for easy differentiation of tissue regions.  However, these images are generated using a single UV wavelength resulting in images with only DNA contrast reminiscent of hematoxylins contribution to standard H&E histology.

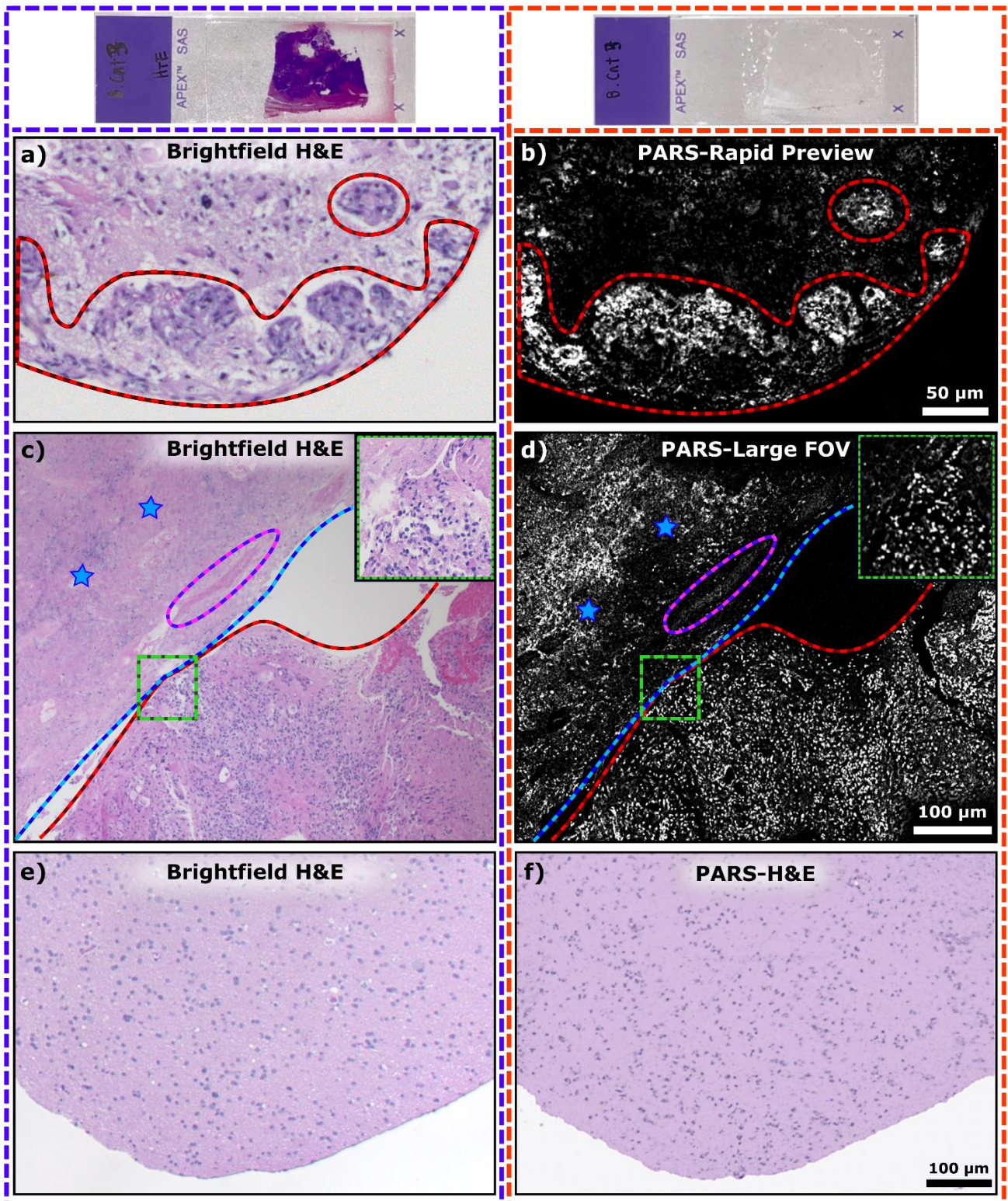

**Figure 2:** PARS H&E images of unstained tissue in comparison to their standard histopathological preparations. The left column of images outlined in purple was taken with a brightfield microscope from an H&E stained slide requiring multiple steps of processing and staining (example shown at top

of column). The right column outlined in orange, was taken with the PARS system from unstained tissue (example shown at the top of column) therefore eliminating the need for sample processing. (a) A conventional bright field H&E image of glioblastoma with solid tumor and microvascular proliferations (red outlines). (b) A rapid acquisition PARS image of the adjacent unstained brain tissue sample, with the region of solid tumoral tissue, and microvascular proliferations (red outlines), as in (a). (c) A standard bright field histopathological H&E image of a glioblastoma sample with a largely necrotic region (blue lines and stars), a thrombotic vessel (purple outline), and a region of solid tumor with microvascular proliferations (red outline). (d) A PARS image acquired in a rapid acquisition mode of the same section of glioblastoma tissue, with the largely necrotic region (blue line and stars) and thrombotic vessel (purple outline), and the solid tumor region with microvascular proliferations (red outline), as in (c). A close-up of tumor cells and microvascular proliferations at the boundary between these regions is shown enclosed in the green boxes. (e) A standard bright field histopathological H&E image of a brain tissue sample with infiltrating tumor cells, adjacent to solid tumor shown in a) – d). (f) A PARS multiwavelength simulated H&E image of the same section of brain tissue.

To produce false-color PARS images with true H&E-like contrast the multiwavelength system was employed. Figure 2e and Figure 2f show a direct comparison between a standard brightfield H&E image of brain tissue with non-malignant gliosis and a PARS H&E-like false-color image of the adjacent section of tissue. The PARS image with nuclear and cytochrome contrast is nearly identical to the adjacent tissue section with standard H&E processing viewed with a transmission brightfield microscope. Subtle clinically relevant features are readily assessed in the PARS image including nuclear and cell morphology, nuclear density and cellular organization. The fidelity of the PARS H&E-like images is further improved by reduction in the field of view. In Figure 3a and 3c, black arrows are used to indicate the location of perivascular oligodendrocytes, a relatively small but clinically relevant deformity observed in this brain tissue. The accompanying PARS H&E image of the adjacent unstained slices are shown in Figures 3b and 3d, respectively. These PARS emulated H&E images are qualitatively identical to the adjacent traditional H&E images, although it should be noted that some minor differences in cellular structure are expected as these images are from adjacent sections 4 µm apart, not from the same slice. Both techniques show easily identifiable perivascular oligodendrocytes, denoted by similar black arrows. This is the first time such features have been identified from unstained tissues in an all optical reflection mode architecture.

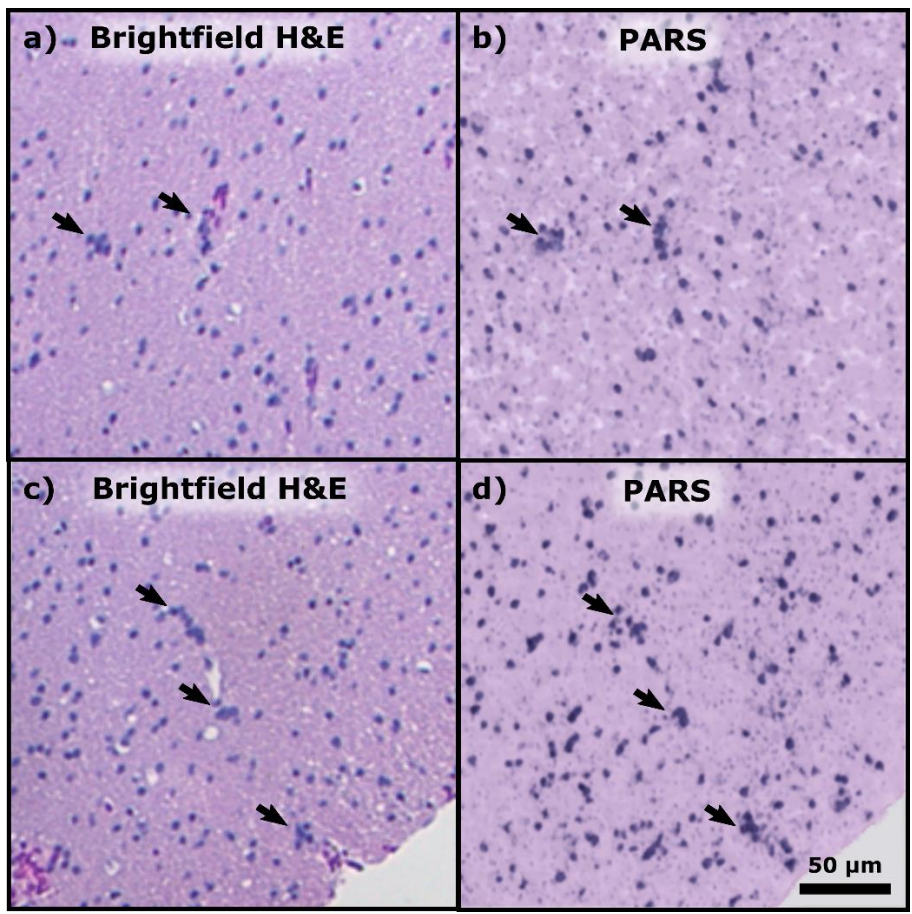

**Figure 3:** Standard brightfield histology, compared to PARS H&E. (a)(c) Small field of view standard H&E processed white matter tissue with gliosis; perivascular oligodendrocytes are indicated by the black arrows. (b)(d) High fidelity PARS emulated H&E images of adjacent tissue sections showing the same clusters of perivascular oligodendrocytes, marked with black arrows.

The PARS architecture explored in this study does not provide depth information in a single interrogation like traditional photoacoustic microscopy. Though a variant of the PARS modality, coherence gated photoacoustic remote sensing microscopy has been proposed, which would provide depth information from a single PARS interrogation[22]. Currently, within thick tissues, PARS provides depth discrimination using optical sectioning. Leveraging the tight axial focus of the UV excitation, PARS can virtually section micron scale layers of cellular morphology. By capturing a series of optically sectioned subsurface images PARS can then recover depth information. Using this strategy, PARS is able to assess subsurface nuclear contrast without modification to the sample. While the current results use sectioned paraffin embedded brain tissues, the all-optical reflection mode architecture lends itself to imaging of thick tissue samples (>1mm). Previously this capability has been shown in thick formalin fixed paraffin embedded breast tissue blocks[17]. Moving forwards volumetric imaging will be explored in appropriate depth resolvable brain tissue samples, such as formalin fixed paraffin embedded tissue blocks, or freshly resected tissues.

## 4    Conclusions

The histological capabilities of the PARS system presented here are well suited to intraoperative guidance of tumor resection surgeries, potentially improving the ability to achieve maximal safe resection. The proposed real time capable PARS microscope has been proven to rapidly assess tissues enabling differentiation of gliosis from healthy tissue, necrotic from living tissues and white from gray matter in unstained brain tissue samples. Furthermore, by using a multiwavelength approach, PARS microscopy generated images of comparable quality to the current gold standard for brain tumor histopathological assessment. Presented here, subtle diagnostic features, oligodendrocytes, have been recovered from unstained human brain tissue for the first time using an all optical reflection mode microscopy system. Notably, the simulated H&E contrast provided by PARS means that if adopted, there may be little requirement to retrain pathologists to interpret a new image type. The provided imaging characteristics can significantly increase the information available to a clinician during neurosurgical resections and can significantly contribute towards an optimal surgical management of malignant brain tumors. To further improve the diagnostic utility of this device, development is focusing on the incorporation of faster repetition rate lasers and additional excitation wavelengths. These improvements will facilitate increased imaging speeds with larger imaging fields, while providing further selective contrast for biomolecules such as lipids and hemoglobin[18]. While the present results demonstrate the capabilities of PARS in the histological analysis of preserved brain tissues, this method will soon be applied to unprocessed freshly resected human brain tissue samples. This work represents a significant step towards the development of a real-time in-situ surgical microscopy system for intraoperative histopathological assessment of the neurosurgical operative field.

**Acknowledgements:**

Authors B. R. Ecclestone and K. Bell contributed equally to this work

This work was supported by Natural Sciences and Engineering Research Council of Canada (grant numbers DGECR-2019-00143, RGPIN2019-06134); Canada Foundation for Innovation (grant number JELF #38000); Mitacs Accelerate (grant number IT13594); University of Waterloo Startup funds; Centre for Bioengineering and Biotechnology (CBB Seed fund); illumiSonics Inc (grant number SRA #083181); New frontiers in research fund – exploration.

**Additional Information**

*Conflict of interest:*

Authors D. Dinakaran, K. Bell, J.R. Mackey and P.H. Reza have financial interests in illumiSonics Inc. IllumiSonics partially supported this work. Authors B.R. Ecclestone, S. Abbasi, F.K.H Landeghem and P. Fieguth declare no competing interests.

*Data Availability:*

All data generated or analyzed during this study are included in this published article.

**Author Contributions:**